# Polarization Invariants and Retrieval of Surface Parameters Using Polarization Measurements in Remote Sensing Applications

**Yuri K. Shestopaloff**[1,*]

[*]*Corresponding author: shes169@yahoo.ca*

Using polarization measurements in remote sensing and optical studies allows retrieving more information. We consider relationship between the reflection coefficients of plane and rough surfaces for linearly polarized waves. Certain polarization properties of reflected waves and polarization invariants, in particular at incident angle of forty five degrees, allow finding amplitude and phase characteristics of reflected waves. Based on this study, we introduce methods for finding dielectric permittivity, temperature and geometrical characteristics of observed surfaces. Experimental results prove that these methods can be used for different practical purposes in technological and remote sensing applications, in a broad range of electromagnetic spectrum.

## 1. Introduction

The vector nature of electromagnetic fields and waves is represented by polarization parameters. Polarization phenomena are closely related to *coherence* of electromagnetic waves. Incoherent electromagnetic emission is non-polarized. On the other hand, it is known that if we receive such a non-polarized signal using a narrow bandwidth filter or a polarization filter for light, then the output will be a signal that has some degree of polarization. So, polarization characteristics depend on the signal parameters, as well as signal interactions with the receiving and emitting devices and propagation media. In this study, we assume that characteristics of emitting and receiving devices are matched, that is, the receiving device obtains the whole spectrum of reflected or emitted signal, so that the signal's polarization properties are not distorted by the receiver.

Polarization properties of electromagnetic signals are often described by parameters that can be considered, from a mathematical perspective, to be independent [1]. On the other hand, the signal polarization properties are defined by characteristics of emitting devices and mediums that interact with this signal. In this sense, the signal's polarization parameters are dependent, although not necessarily in a direct way. For instance, reflection coefficients on parallel and perpendicular polarizations both depend on the dielectric permittivity of reflecting medium. Once exposed, such inherent relationships can enhance our understanding of the nature of electromagnetic signals and their interaction with media. In particular, they allow us to extract more information from polarization measurements, and increase the accuracy of their interpretation. However, in many instances, the relationship of polarization parameters and characteristics of studied objects is not straightforward. This is why the interpretation of polarization measurements can be ambiguous. Given this, the existence of special observation conditions, when polarization parameters and characteristics of studied objects have simple physical, and accordingly mathematical, relationships, can be beneficial for practical applications. Such polarization invariants are discussed in this article.

First, we present the results of a theoretical study of polarization properties of linearly polarized electromagnetic waves reflected from flat and rough surfaces. Based on this study, we introduce methods for determining characteristics of reflected and refracted signals, and methods for finding dielectric permittivity, temperature and geometric characteristics of observed objects. We also suggest a calibration method for determining the noise level. Then, in a separate section, we present the results of experimental studies of the proposed methods and discuss their practical applications.

## 2. Relationship of reflection coefficients and emissivities of linearly polarized waves. Retrieving a medium's parameters

Let us consider a flat surface. In our study, we use the *direct* relationship between the reflection coefficient on parallel polarization and the reflection coefficient on perpendicular polarization, at the *same* arbitrary incident angle. In this, we follow the approaches suggested in [2,3]. Let us consider the complex reflection coefficients $R_p$ and $R_s$, which correspond, respectively, to parallel and perpendicular polarizations. The magnetic permittivities of both media are assumed to be equal to one.

$$R_s = \frac{\sqrt{\varepsilon - \sin^2 \alpha} - \cos \alpha}{\sqrt{\varepsilon - \sin^2 \alpha} + \cos \alpha} . \tag{1}$$

$$R_p = \frac{\varepsilon \cos \alpha - \sqrt{\varepsilon - \sin^2 \alpha}}{\varepsilon \cos \alpha + \sqrt{\varepsilon - \sin^2 \alpha}} . \tag{2}$$

Here, $\alpha$ is the incidence angle (angle between the normal to the surface and the direction of observation) and $\varepsilon$ is the dimensionless dielectric permittivity, which is *complex-valued*. Therefore, in general, the reflection coefficients incorporate phase information. Formula (2) defines the Fresnel reflection coefficient on parallel polarization, while formula (1) defines the Fresnel reflection coefficient on perpendicular polarization, taken with the opposite sign. The reason for the sign change is this. At normal angle of incidence $\alpha = 0$, reflection coefficients have to be equal, because the observation conditions for the *reflected* waves are *identical* in this case, and measurements on perpendicular and parallel polarizations are *indistinguishable*. Therefore, the complex values of respective reflection coefficients should be equal. This is possible only if the reflection coefficients are presented in the form of (1) and (2). Indeed, in this case, we have for the normal angle of incidence $R_s = R_p = \left(\sqrt{\varepsilon} - 1\right)/\left(\sqrt{\varepsilon} + 1\right)$. If we would like to use *incident* waves as the reference point for the phases of reflection coefficients, then we would obtain $R_s = -R_p$, which is incorrect, because, as we found out, observations at normal angle of incidence on both polarizations are identical, and consequently reflection coefficients should be the same. The aforementioned consideration is by no means a principal one, because, essentially, we are talking about what reference point is to be taken to calculate the phase shift between the complex reflection coefficients. So, if one decides to use (1) with the opposite sign, that is in a traditional form that computes the phase shift relative to *incident* waves, then all results that we obtain in this article remain valid if one substitutes $R_s$ by $(-R_s)$.

Note that (1) and (2) also represent the general case of reflection from the flat boundary between mediums with arbitrary dielectric permittivities $\varepsilon_1$ and $\varepsilon_2$, where the index “2” corresponds to reflecting medium. In this case, we should assume in (1) and (2) that $\varepsilon = \varepsilon_2 / \varepsilon_1$. So, all results that we obtain below are also valid for the case when the incident wave comes from a medium whose dielectric permittivity differs from one. Certainly, the dielectric permittivity of both media can be complex.

There is an interesting reflection property of a boundary between media that have the same dielectric permittivity but different magnetic permittivity. In this case, the refection coefficients on both polarizations are the same and do not depend on observation angle. In particular, this means that the reflection coefficients of a surface with large irregularities (whose linear size is substantially greater than the wavelength) and a flat surface are the same. This effect might have a practical value. More details are in the Appendix.

Using the value of dielectric permittivity from (1) and substituting it into (2), we find that

$$R_p = \frac{R_s^2 + R_s \cos 2\alpha}{1 + R_s \cos 2\alpha}. \qquad (3)$$

(The derivation is in the Appendix, formulas (A1) – (A3).)

Note that since the equations (1) – (3) are complex-valued, (3) includes the relationship between the absolute values and phases of reflection coefficients. Let us assume that $R_s = |R_s|\exp(i\varphi_s)$, and $R_p = |R_p|\exp(i\varphi_p)$. Substituting these values into (3) and equating the real and imaginary parts, we obtain the following system of equations.

$$|R_p|\sin\varphi_p + |R_s||R_p|\cos 2\alpha \sin(\varphi_s + \varphi_p) = |R_s|^2 \sin(2\varphi_s) + |R_s|\cos 2\alpha \sin\varphi_s . \quad (4)$$

$$|R_p|\cos\varphi_p + |R_s||R_p|\cos 2\alpha \cos(\varphi_s + \varphi_p) = |R_s|^2 \cos(2\varphi_s) + |R_s|\cos 2\alpha \cos\varphi_s . \quad (5)$$

Using (4), (5), we can find phases $\varphi_s$ and $\varphi_p$ through the absolute values of reflection coefficients. In order to find the solution, we may square both left and right sides of (4) and (5) and sum up the appropriate sides of these equations. After transformations, we obtain the following.

$$\cos\varphi_s = \frac{|R_s|^4 + |R_s|^2 \cos^2 2\alpha - |R_p|^2 (1 + |R_s|^2 \cos^2 2\alpha)}{2|R_s|\cos 2\alpha (|R_p|^2 - |R_s|^2)} . \quad (6)$$

Now, using (4) or (5), we can find the value of $\varphi_p$ by solving the appropriate equation with respect to either $\sin\varphi_p$ or $\tan(\varphi_p/2)$, depending on the chosen transformation.

In practice, it is usually easier to measure the appropriate emissivities $\chi_p = 1 - |R_p|^2$ and $\chi_s = 1 - |R_s|^2$ as opposed to complex reflection coefficients with phase shifts. This way, we measure characteristics relevant to *energy*, while the phase information is destroyed. (Here and below we use the definition of emissivity

as “the ratio of the radiation emitted by a surface to the radiation emitted by a perfect blackbody at the same temperature” from [4]. However, one should understand that radiation is emitted from the surface layer that has finite thickness.) Using the proposed approach, we can restore the phases of reflection coefficients when we measure only the energy characteristics of signals. In the case of $\alpha = 45^0$, the phase values in (6) are undefined. However, it follows from (3) that the following relationships hold true regardless of the absolute values of reflection coefficients and, consequently, the dielectric permittivity.

$$R_p = R_s^2 . \qquad (7)$$

$$\varphi_p = 2\varphi_s . \qquad (8)$$

First, this effect was discovered in [5]. In [6], it was obtained based on different considerations as a limit of certain expression. Physically, the relationships (7) and (8) are due to the geometry of interaction of electrical dipoles (oscillating electrical charges) near the medium’s surface with vector electromagnetic field of incident waves. In order to understand the effect, the analogy with Brewster’s angle can be used. Radiation diagrams for the Brewster angle and when the angle of incidence is $45^0$ are shown in Fig. 1.

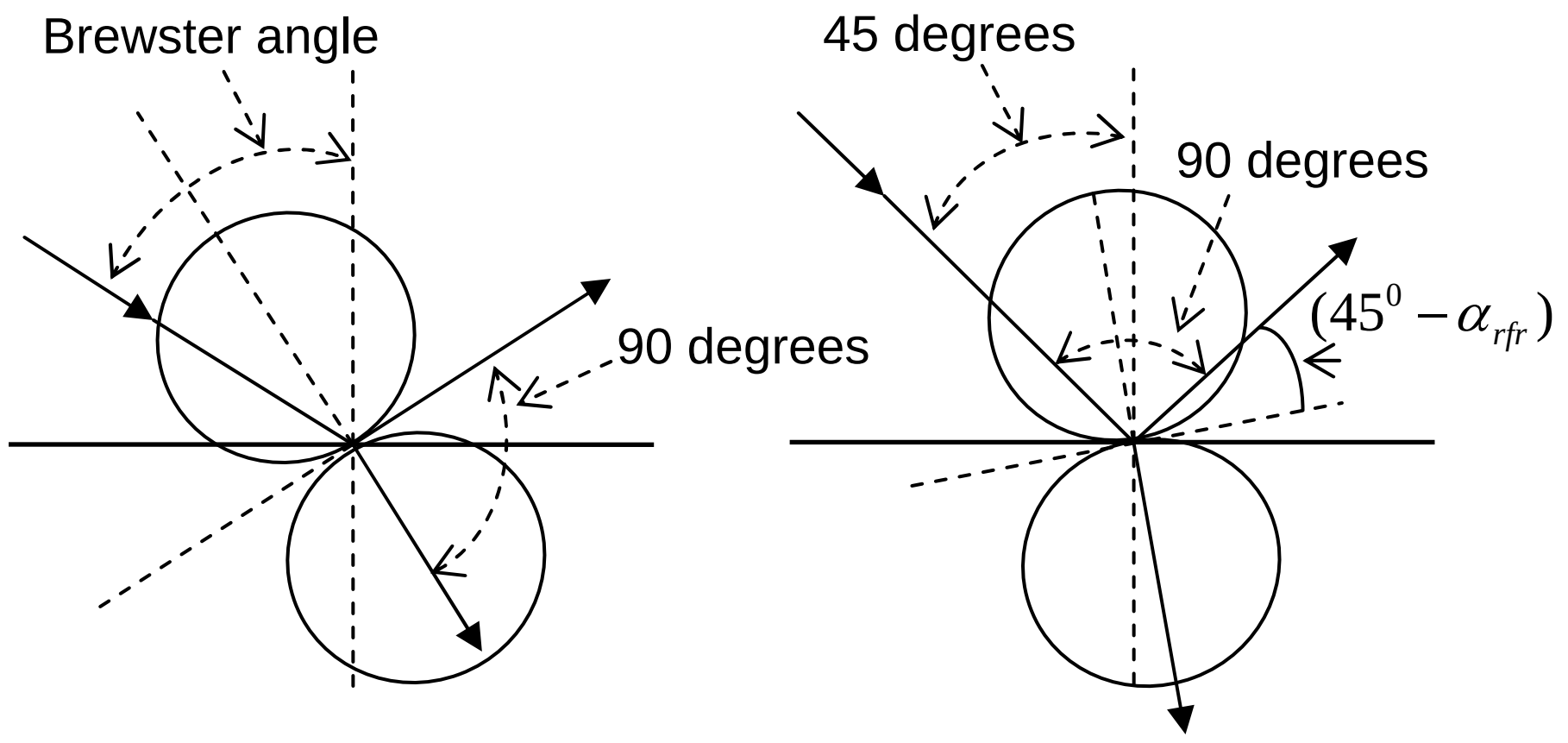

Fig. 1. Radiation diagrams (circles) for the Brewster's angle (left) and $45^0$ angle (right).

Theoretically, the reflected wave at parallel polarization is absent at Brewster's angle, because electrical charges of the medium oscillate perpendicular to the direction of the refracted wave, which is also perpendicular to the direction of the reflected wave. The radiation diagram of these oscillating charges is such that there is no emission in the direction of the reflected wave at parallel polarization. However, the reflected wave is actually the result of interaction of radiation of electrical charges of the media, that are excited by the incident electromagnetic wave, with the incident electromagnetic wave. This process happens in a layer of the medium with *finite* thickness. Consequently, in practice, the reflected signal with parallel polarization never becomes exactly zero at Brewster's angle. In work [7], the author says: "Studies of the light reflected at Brewster's angle showed that there are small deviations from the values predicted by Fresnel's formulas. It turns out that there is no such incident angle when the intensity of the reflected wave at the parallel polarization becomes zero and the electrical vector of the reflected wave oscillates perpendicular to the plane of incidence." The author explains that this effect occurs because "reflection and refraction happen not on a mathematical surface boundary, but in a thin transactional layer between different media".

In case of incident angle of $45^0$, the directions of incident and reflected waves are perpendicular, as it is shown in Fig. 1. The direction of the refracted wave (let us denote it by angle $\alpha_{rfr}$ counted from the normal direction) differs from perpendicular to the reflected wave by $(45^0 - \alpha_{rfr})$. So, unlike in the case of Brewster's angle, the radiation diagram of dipoles is oriented in such a way that the emitted signal is not zero at parallel polarization in the direction of reflected wave. At the same time, the existence of simple relationship (7) means that there is a certain *physical* connection between the parallel polarized radiation emitted by dipoles at angle $(45^0 - \alpha_{rfr})$ relative to the axis of symmetry of their radiation diagram, and the radiation emitted by dipoles that are excited by the electric field of a perpendicularly polarized wave. The explanation of the details of this effect requires further study.

What would be the use of this effect? Electrical charges on the surface should not radiate in the same way as electrical charges that are deeper in the medium. The orientation of their radiation pattern might be different. So, actually, we may benefit from this effect at least in two ways. The detailed study of certain electrodynamic symmetry, which (7) and (8) present, should give us a better understanding of the nature of electromagnetic waves, because Fresnel's formulas follow from Maxwell's equations. On the other hand, the experimentally measured deviations from (7) and (8) would allow us to evaluate effects caused by the finite thickness of the boundary layer, in which the processes of reflection and refraction actually take place, and so allow us to better understand the properties of this layer.

Another useful consequence of (1) and (2) is that we can express the complex-valued dielectric permittivity through the complex-valued reflection coefficients (derivation is in Appendix, formula (A4)) as follows.

$$\varepsilon = \frac{(1+R_p)(1+R_s)}{(1-R_p)(1-R_s)}. \tag{9}$$

We can validate (9) as follows. If we assume that $\alpha = 0$ in (1) and (2), then we can find $\varepsilon = (1+R_s)^2/(1-R_s)^2$ and $\varepsilon = (1+R_p)^2/(1-R_p)^2$ respectively. At normal angle of incidence, reflection coefficients in (9) are equal, and so formula (9) produces the same values, which confirms its correctness.

We can also find the dielectric permittivity separately from (1) or (2) when we know the observation angle and the appropriate reflection coefficient. However, in the case of (9), we only have to measure the reflection coefficients, and we do not need to measure the observation angle. This is an advantageous feature, because accurately measuring observation angles often presents a challenge, especially in remote sensing applications. Note that we can use several measurements at different angles with (9). Then, if the measurement errors are not correlated, their average value will produce more accurate results. In fact, in many instances, several such measurements, aided by some additional information, let us say about surface roughness, can allow the discovery of systematic errors as well, thus also reducing the total error.

An interesting thing about (9) is that by combining (4) – (6) and (9) we can find the complex-valued dielectric permittivity without knowing the phase structure of reflection coefficients, only their absolute values. For instance, we can measure emissivites $\chi_p = 1 - |R_p|^2$ and $\chi_s = 1 - |R_s|^2$ (which do not contain phase information) and find the phases of reflection coefficients $\varphi_s$ and $\varphi_p$ using (4) – (6). Then, (9) can be rewritten as follows.

$$\varepsilon = \frac{(1+\sqrt{1-\chi_p}(\cos\varphi_p + i\sin\varphi_p))(1+\sqrt{1-\chi_s}(\cos\varphi_s + i\sin\varphi_s))}{(1-\sqrt{1-\chi_p}(\cos\varphi_p + i\sin\varphi_p))(1-\sqrt{1-\chi_s}(\cos\varphi_s + i\sin\varphi_s))}. \qquad (10)$$

## 3. Polarization invariant

At observation angle of $45^0$, equation (3) can be viewed as a polarization invariant. For complex reflection coefficients, we can rewrite it as follows.

$$\frac{R_p}{R_s^2} = 1. \qquad (11)$$

If we use emissivities for the respective polarizations, then (11) can be presented in the following form.

$$\frac{\chi_s^2}{2\chi_s - \chi_p} = 1. \qquad (12)$$

The general form of polarization invariant (11) for an arbitrary angle of incidence is as follows.

$$\frac{R_p(1+R_s\cos 2\alpha)}{R_s^2+R_s\cos 2\alpha}=1 \cdot \qquad (13)$$

An equation that is similar to (13), but written in terms of emissivities, can be derived from (5) by performing the appropriate transformations. Using $\cos\varphi_s$ from (6), we obtain the following.

$$\chi_p=\frac{2\chi_s(1+\cos 2\alpha\cos\varphi_s\sqrt{1-\chi_s})-\chi_s^2}{1+\cos^2 2\alpha(1-\chi_s)+2\cos 2\alpha\cos\varphi_s\sqrt{1-\chi_s}} \,. \qquad (14)$$

Transforming (14), we find that the polarization invariant for an arbitrary angle of incidence is as follows.

$$\frac{\chi_s^2}{2\chi_s+2\cos 2\alpha\cos\varphi_s\sqrt{1-\chi_s}(\chi_s-\chi_p)-\chi_p\cos^2 2\alpha(1-\chi_s)-\chi_p}=1 \cdot \qquad (15)$$

If the imaginary part of the dielectric permittivity is small and can be neglected, then we can assume $\varphi_s=0$.

## 4. Finding temperature using polarization measurements

The forms of polarization invariant (12) and (15) are more appropriate when we consider electromagnetic radiation emitted by the body itself, for instance, infrared or microwave radiation [8]. In the case of polarization measurements, we can define the respective emissivities as follows: $\chi_s=\frac{T_s}{T}$, $\chi_p=\frac{T_p}{T}$, where $T_s$ and $T_p$ are the brightness temperatures on perpendicular and parallel polarizations; $T$ is the thermodynamic temperature of a remote object of interest. (This follows from the previously mentioned definition of emissivity from [4].) Substituting these values into (12), we can find the thermodynamic temperature of a remote object by using measurements on two polarizations as follows.

$$T = \frac{T_s^2}{2T_s - T_p}. \tag{16}$$

The physical importance of (16) is that it ties polarization measurements of electromagnetic emission in *any range* of electromagnetic spectrum (to which Fresnel formulas and Maxwell's equations can be applied), to thermodynamic temperature. This is a consequence of the continuity of the spectrum of electromagnetic emission emitted by bodies whose temperature is not zero (Planck spectrum). For instance, it is possible to find the temperature of the Earth's surface using satellite or airborne measurements of the Earth's microwave radiation.

Using (15), we can derive a formula similar to (16) for an arbitrary observation angle. In order to do that, we should multiply the numerator and denominator in (15) by the temperature *T*. Let us also denote $x = \sqrt{1-\chi_s}$. Then, $x^2 = 1-\chi_s$ and $\chi_s = 1 - x^2$. Substituting these values into (15) and doing appropriate transformations, we obtain the following equation.

$$\frac{T_s(1-x^2)}{2T_s + 2x\cos 2\alpha \cos\varphi_s (T_s - T_p) - x^2 T_p \cos^2 2\alpha - T_p} = 1 \quad . \tag{17}$$

Transforming (17), we obtain the following quadratic equation.

$$x^2(T_p \cos^2 2\alpha - T_s) - 2\cos 2\alpha \cos\varphi_s (T_p - T_s)x + (T_p - T_s) = 0. \tag{18}$$

Generally, it has two roots. Positive sign and a value of less than 1 is a good indicator of which solution is the right one. If both roots satisfy to the conditions, then the choice should be based on input data. Once we know the correct solution *x*, the temperature can be found as $T = T_s / (1 - x^2)$. If $\varphi_s = 0$, then solution of (18) becomes

$$x_{1,2} = \frac{\cos 2\alpha (T_p - T_s) \pm \sin 2\alpha (T_p T_s - T_s^2)^{0.5}}{T_p \cos^2 2\alpha - T_s}. \tag{19}$$

For $\alpha = 45^0$, (19) transforms to (16), which serves as a validation of (19). In practical applications, the value of $\varphi_s$ may be unknown. However, in many instances, the imaginary part of the dielectric permittivity is very small. So, for such observations, we can assume $\varphi_s = 0$.

## 5. Polarization invariant as an unambiguous criterion for surface roughness

Recall that the polarization invariants (11), (13), (15) have been derived for a flat surface. When surface roughness changes, the value of $R_p / R_s^2$ in (11) will change accordingly. We can then define a quantitative characteristic of the surface roughness *S* as follows.

$$S = \frac{R_s^2}{R_p}. \tag{20}$$

Note that *S*, in general, can be complex-valued. The parameter *S* (let us call it the *roughness coefficient*) is sensitive to all types of surface irregularities. It is especially sensitive when the linear size of roughness is greater than the wavelength. To understand the influence of surface irregularities on the roughness coefficient, we used models suggested in [8-10] for computing reflection coefficients. Here, the rough surface was modeled by a combination of different geometrical forms at different observation angles. In the range of dielectric permittivity from 2.5 to 70, an increase in the average slope of the surface's irregularities leads to a monotonic decrease of reflection coefficient on perpendicular polarization and monotonic increase of reflection coefficient on parallel polarization, when compared to a medium with the same dielectric permittivity but a flat surface. So, in this case, the roughness coefficient (20) monotonically decreases starting from a value of one, which corresponds to a flat surface.

Instead of reflection coefficients, we can also use emissivities to introduce a similar roughness coefficient. For instance, for the angle $\alpha = 45^o$, the roughness coefficient $S_\chi$ is as follows (note that it has a maximum value of one for a flat surface).

$$S_\chi = \frac{\chi_s^2}{2\chi_s - \chi_p}. \qquad (21)$$

In this case, $S_\chi$ is real-valued, not complex-valued as is the case with (20). Note that when the surface is flat, both (20) and (21) are real. The imaginary part of *S* in (20) can appear only for non-flat surface.

The appropriate roughness coefficient for an arbitrary angle of observation is as follows.

$$S_\chi = \frac{\chi_s^2}{2\chi_s + 2\cos 2\alpha \cos\varphi_s \sqrt{1-\chi_s}\,(\chi_s - \chi_p) - \chi_p \cos^2 2\alpha (1-\chi_s) - \chi_p}. \qquad (22)$$

Fig. 2 illustrates the dependence (22) graphically for two values of dielectric permittivity.

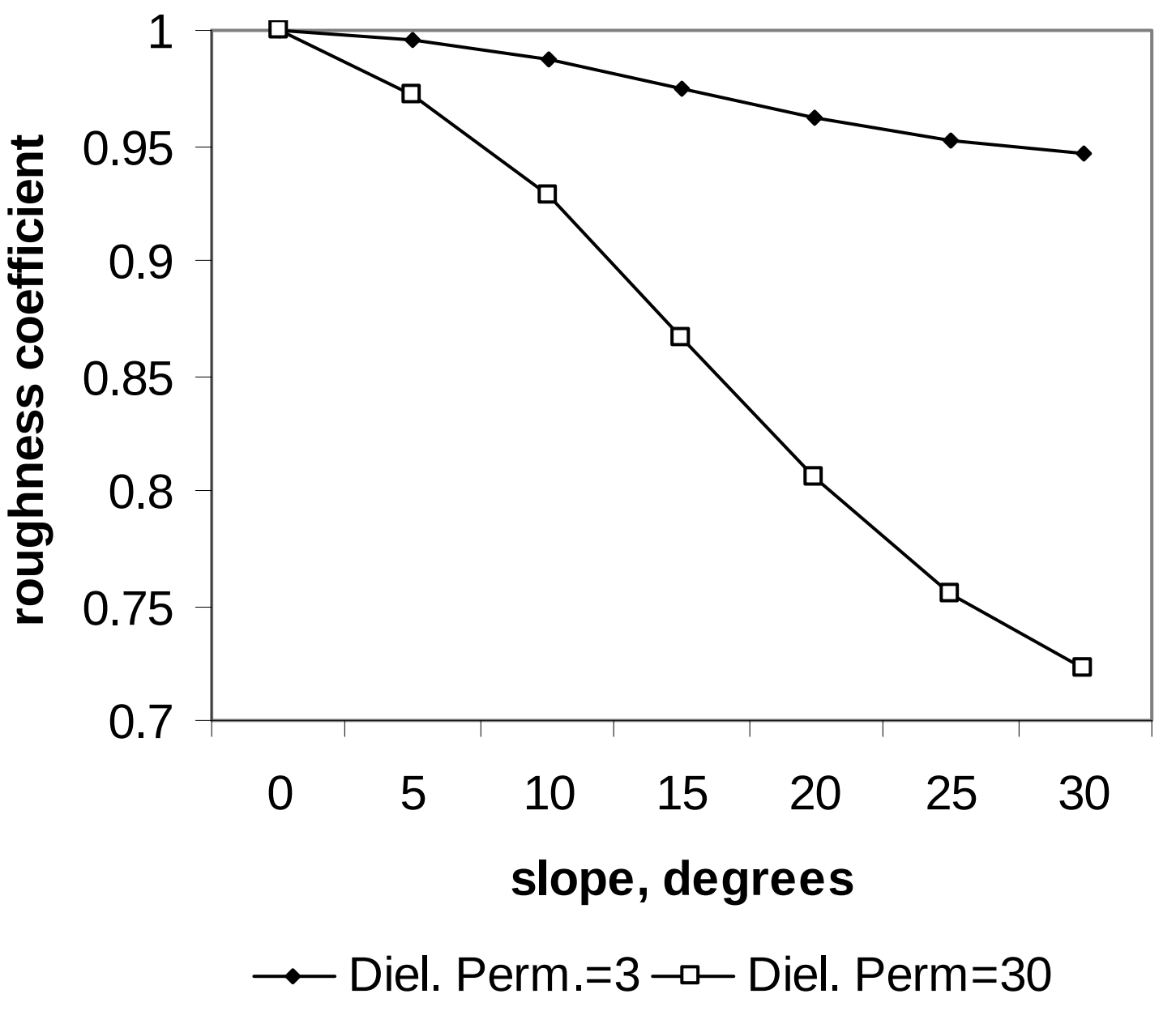

Fig. 2. Dependence of roughness coefficient on the average slope of a rough surface modeled by an assembly of cones, at an observation angle of $45^0$.

## 6. Using polarization invariant for calibration

In practical applications, received signals are often accompanied by some background radiation, which can come from external sources such as relic background radiation or cloud emission in microwave measurements, noise of electronic receiving device, or white noise in optical measurements. The good thing is that calibration procedures, to compensate for noise, can be built into the device itself based on the polarization invariant.

Let us assume that there is no external noise, and the device's, or internal, noise level is *N*. Let $T_{sr}$ and $T_{pr}$ be non-calibrated measurements of brightness temperatures on perpendicular and parallel polarizations respectively, so that the calibrated brightness temperatures are $T_s = (T_{sr} - N)$ and $T_p = (T_{pr} - N)$, and the thermodynamic temperature of emitting object is *T*. For instance, in the case of observation angle of $45^0$, we can substitute these values into (16) and obtain a quadratic equation with respect to *N*. Solving this equation, we obtain the following two roots. (Which of the two roots has a practical meaning is determined by the device's specifics.)

$$N = 0.5\left\{(2T_{sr} - S_\chi T) \pm \left[S_\chi^2 T^2 + 4S_\chi T(T_{sr} - T_{pr})\right]^{1/2}\right\}. \qquad (23)$$

The advantage of this method is that it does not require us to know the value of the dielectric permittivity and emissivity of the observed object in order to estimate the noise level. Certainly, we can use the same approach to estimate the value of external noise. If the nature of external and internal noises is the same, then external noise can be included into the overall noise defined by (23). If the internal or external noise depends on the polarization, we still can use this approach. However, we have to add a second measurement at a different angle. This way, we obtain two mathematically independent equations, and accordingly can find two unknown variables, which are the noise for each polarization channel. For instance, when phase shift $\varphi_s = 0$, the appropriate second equation, based on (19), is as follows.

$$\begin{aligned}&\left(1-(T_s - N_s)/T\right)^{0.5} = \\ &= \frac{\cos 2\alpha(T_p - T_s - N_p + N_s) \pm \sin 2\alpha[(T_p - N_p)(T_s - N_s) - (T_s - N_s)^2]^{0.5}}{(T_p - N_p)\cos^2 2\alpha - (T_s - N_s)}\ .\end{aligned} \qquad (24)$$

Here, *N* represents the value of noise in the appropriate polarization channel.

## 7. Experimental observations and discussion

## *Temperature determination*

Proposed methods for remote temperature determination on the basis of polarization measurements were tested experimentally. The measurements were performed outdoors. In the case of 3.4 cm wavelength, the antenna's projector zone was used for observations. For the 3.4 cm and 1.25 cm wavelengths, an adjustment $T_b$ was made for the background radiation coming from the sky. The value of radiation from the sky $T_{sky}$ was measured by the same radiometer; it could be up to 15 K in cloudy weather. Then, using the values of dielectric permittivities from [11,12] or the ones found in our previous experimental studies, we evaluated the background radiation as follows: $T_b = T_{sky}|R|^2$, where $R$ corresponds to the reflection coefficients (1) or (2) for the appropriate polarization at the angle of observation we used. Finally, the adjustment $T_b$ was subtracted from the measured brightness temperature. Measurement at 2.2 cm wavelength was done from a tower at a distance at which the antenna's radiation diagram was fully formed. In all cases, the receivers' error was about 0.7 K or less.

Table 1 presents the average absolute error in our experiments on temperature determination by microwave radiometers for different surfaces and different wavelengths.

Table 1. Accuracy of temperature determination ($^0C$) observed in experiments.

| Angle, degrees / Surface (wavelength, cm) | 30 | 45 | 60 |
|---|---|---|---|
| fresh water (3.4) | 3.7 | 2.5 | 2.6 |
| salt water (3.4) | 4.2 | 3.1 | 4.2 |
| dry sand (3.4) | 5.9 | 4.4 | 3.8 |
| wet sand (3.4) | 4.0 | 3.3 | 3.6 |
| dry fresh snow (2.2) | 33 | 27 | 29 |
| dense snow (2.2) | 12 | 9 | 11 |
| plain wood surface (1.25) | 1.8 | 1.6 | 1.6 |

We can evaluate the accuracy $\delta T$ of temperature determination by differentiating (16). Assuming that the measurement errors of brightness temperatures are accordingly $\delta T_s$ and $\delta T_p$, we obtain the following.

$$\delta T = \frac{2T_s^2 \delta T_s + T_s (T_s \delta T_p - 2T_p \delta T_s)}{(2T_s - T_p)^2} . \quad (25)$$

In many instances, we can assume that $\delta T_s \approx \delta T_p$. Denoting $\delta T_{bt} = \delta T_p = \delta T_s$, we can rewrite (25) as follows.

$$\delta T \approx \delta T_{bt} \frac{T_s (T_s - 2T_p) + 2T_s^2}{(2T_s - T_p)^2} . \quad (26)$$

We can see from (26) that when the brightness temperatures are close, that is $T_s \approx T_p$, which was the case with fresh dry snow and dry sand, we have $\delta T \approx \delta T_{bt}$. Given this, it is not clear why the error in case of fresh snow is so large, even we take into account the influence of small surface roughness (when the linear size of surface roughness is less than the wavelength). The most likely reason is that the snow layer was not electrophysically uniform. The thickness of the electromagnetic "skin-layer" was several tens of wavelengths, because of the low snow density and low snow conductivity. So, in fact, the snow we were dealing with was a *multilayer* medium, composed of several layers that fell at different times and were subjected to changes under the influence of temperature variations, sunlight and wind. These factors could have caused the large observed error. Otherwise, the estimates based on (16) and (18) correlated with experimental observations relatively well. In general, the higher the dimensionless dielectric permittivity, the more accurate temperature measurements are. Except for the case of dry snow, the main sources of errors are background radiation and influence of small scale surface roughness. Both factors can potentially be taken into account based on a priori information about the surface and observation conditions. The presence of systematic errors can also indicate the existence of some constant factors, e.g. calibration errors.

Infrared measurements are widely used for temperature determination. In some instances, when the observed surfaces are smooth and the observation angle is not normal, the accuracy of temperature measurements can potentially be increased by using polarized infrared measurements.

## *Measurement of roughness coefficient*

We studied experimentally the possibility of using the introduced roughness coefficient for practical evaluation of average slopes of rough surfaces. Observations were done with 3.4 cm radiometer in the antenna's projector zone. We used sand to model a rough surface consisting of an assembly of cones, whose linear length exceeded the radiometer's wavelength by several times. The sand composition was silica; it had a typical saturated sandy color, average size grain of the order of tenth of a millimeter, with the value of the dielectric permittivity of dry sand of 3.06. For the experiments, the sand was moderately moistened. We repeated the experiment four times for different values of dielectric permittivity (adding water and mixing the sand). The dielectric permittivity $\varepsilon$ of wet sand was in the range 5.5 – 7.4, which was found using the measured emissivities of sand with a flat surface.  The moisture content of wet sand was not required since we measured the dielectric permittivity directly.

We used formulas (9) and (10) for calculation of dielectric permittivity both for the slope and nadir observations. The imaginary part of dielectric permittivity was very small, and we considered the dielectric permittivity to be real-valued. In order to verify the accuracy, we also used the average value of emissivity found from nadir observations on two orthogonal polarizations. A dependence of the experimentally determined roughness coefficient on the average slope versus the theoretical curve is shown in Fig. 3. We can see that experimental results closely follow the theoretical curve.

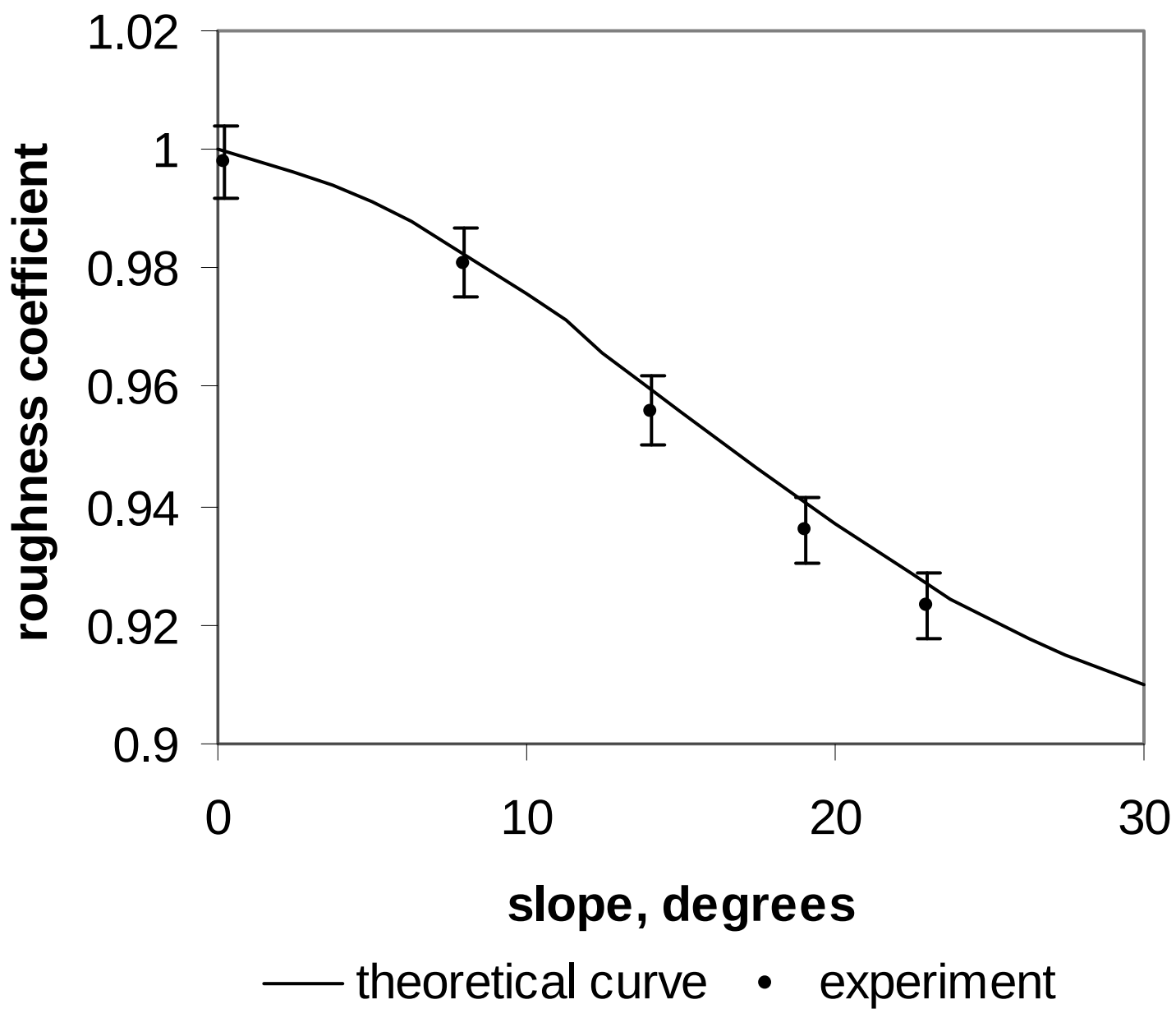

Fig. 3. Roughness coefficient of the rough sand surface measured by 3.4 cm radiometer at $45^0$ observation angle versus theoretical curve. Dimensionless dielectric permittivity is 7.4; roughness is modeled by sand cones. Error bars correspond to one standard deviation.

The measurement error was estimated by differentiating (21).

$$\delta S_\chi = \frac{2\chi_s^2 \delta\chi_s + \chi_s(\chi_s \delta\chi_p - 2\chi_p \delta\chi_s)}{(2\chi_s - \chi_p)^2}. \tag{27}$$

where $\delta\chi_p$ and $\delta\chi_s$ are the measurement errors of the appropriate emissivities. Experimental results in Fig. 3 are within the range of possible errors defined by (27). There is a clear indication of systematic error in Fig. 3, which was satisfactorily explained by the cooperative influence of small scale roughness and by the systematic measurement error of angle of observation. (The influence of small scale roughness does not depend much on the slope of large scale roughness, as we found from previous experiments. The same result was reported in theoretical and experimental studies of surfaces with small scale roughness at different angles of observation and different wavelengths [8,9,13].)

## *Interpretation of satellite observations*

We analyzed data from satellite "Kosmos-1151" acquired by a polarization radiometer at 3.4 cm wavelength (two channels on parallel and perpendicular polarizations, and one nadir channel). The angle of observation was about $60^0$. The time of observation was February 2, 1980, 8 AM Greenwich time. The satellite's position and orientation data were available, so that the location of antenna spot and observation angle were calculated accurately. The linear size of an antenna spot is on the order of ten kilometers. Other available nadir channels of radiometer at 0.8, 1.35 and 8.5 cm were used to assess additional information, such as temperature variations, atmosphere parameters, and also to verify values of dielectric permittivity. The dielectric permittivity can be evaluated using both nadir and oblique angles of observation. We used both approaches for comparison purposes. Below, for simplicity, we discuss nadir measurements.

The errors were estimated using differentials of formulas for evaluation of appropriate parameters, similar to (25), (27). Based on available data, we found that the relative error of evaluation of dielectric permittivity was about $\delta\varepsilon/\varepsilon \approx 10\delta\chi_n$, where $\delta\chi_n$ is the relative error of valuation of emissivity at nadir observation. This inaccuracy was produced by errors of temperature determination (3 K), radiometer noise (1 K), calibration errors (1 K). The errors are independent, so we used their square average, which in our case was $\delta\chi_n \approx 0.018$, so that $\delta\varepsilon / \varepsilon \approx 0.18$.

The error of determination of steepness of the surface depends on the accuracy of valuation of emissivity, error of valuation of dielectric permittivity and imperfections of our rough surface model. Taking into account these factors, we found that for the average values of steepness of $10^0$, $20^0$, $30^0$ the errors were accordingly $2.6^0$, $1.8^0$, $1^0$.

The antenna's spot trajectory went approximately over 6-th meridian of western longitude and between 16 and 29 degrees of northern latitude, which included Sahara desert and the Atlas Mountains' foothills in Northern Africa. The results of computing the average slope from experimental data, in particular from the roughness coefficient, are presented in Fig. 4. Note that the roughness parameter that is based on polarization measurements is sensitive to *all* surface irregularities whose linear scale is greater than several wavelengths. So, large scale geomorphologic and geographic characteristics, which are of the order of a hundred meters and more (let us call them "relief unevenness"), are also included into this integral surface roughness defined by polarization measurements.

The roughness coefficient allows the introduction of a new characteristic of earth surfaces, namely the roughness of small scale surface irregularities (let us call them "surface roughness"), which are not assessed by geographic methods. This is because, in many instances, the contributions of different scale relief unevenness and surface roughness into the integral roughness coefficient can be separated based on preliminary information and additional measurements. This further increases the information capabilities of polarization measurements with regard to evaluation of surface geometry.

Fig. 5 presents the elevation profile for the same trajectory in 90-meter resolution [14], averaged over the antenna spot using values of elevation in equally distanced 30 points, while Fig. 6 shows the average slope over the antenna spot. We used the slope data with one square kilometer resolution [15], so that the average slope was calculated over the larger area corresponding to the antenna spot. We used data with discretization of one degree of slope, so that even if the actual values of slopes were not zero, but less than one degree, they were assumed to be equal to zero. Similarly, all slopes between one and two degrees were assumed to be equal to one degree, and so on. So, the calculated values have bias toward smaller than actual values of average slopes. This is why the calculated average value of the slope is zero through a large part of the trajectory. Within the antenna spot, there were simultaneously areas with different slopes, so that the produced average values of the slope could have a fractional value. The average error was calculated based on the assumption that the slope values are uniformly distributed within each interval of one degree.

Generally, terrain that is rougher in a geological sense is also rougher on a smaller scale. For instance, a mountain area has stones on the surface, while flat valleys usually do not. Similarly, higher elevation correlates with rougher terrain. These relationships are clearly observed in the presented figures.

We can see that our evaluation of the average surface slope correlates well both with elevation data and changes of average slope. Besides these comparisons, we also did a visual analysis of detailed optical images with resolutions of a few tens of meters (available on public website “Google Maps”). Qualitatively, the results are in very good agreement with optical data, both for the surface roughness and dielectric permittivity. Average steepness, as an integral characteristic, also correlates well with specifics of certain geographical objects, such as plateaus, ridges of sand dunes, salt pans, foothills, etc; see comments below.

Although not presented in the article, we also used information from other sources for comparison, such as geomorphologic data [16], weather conditions, etc, which can be translated into values of dimensionless dielectric permittivity and average slope. This prior geological and geomorphologic information also corresponds well to the results we obtained from interpretation of satellite measurements.

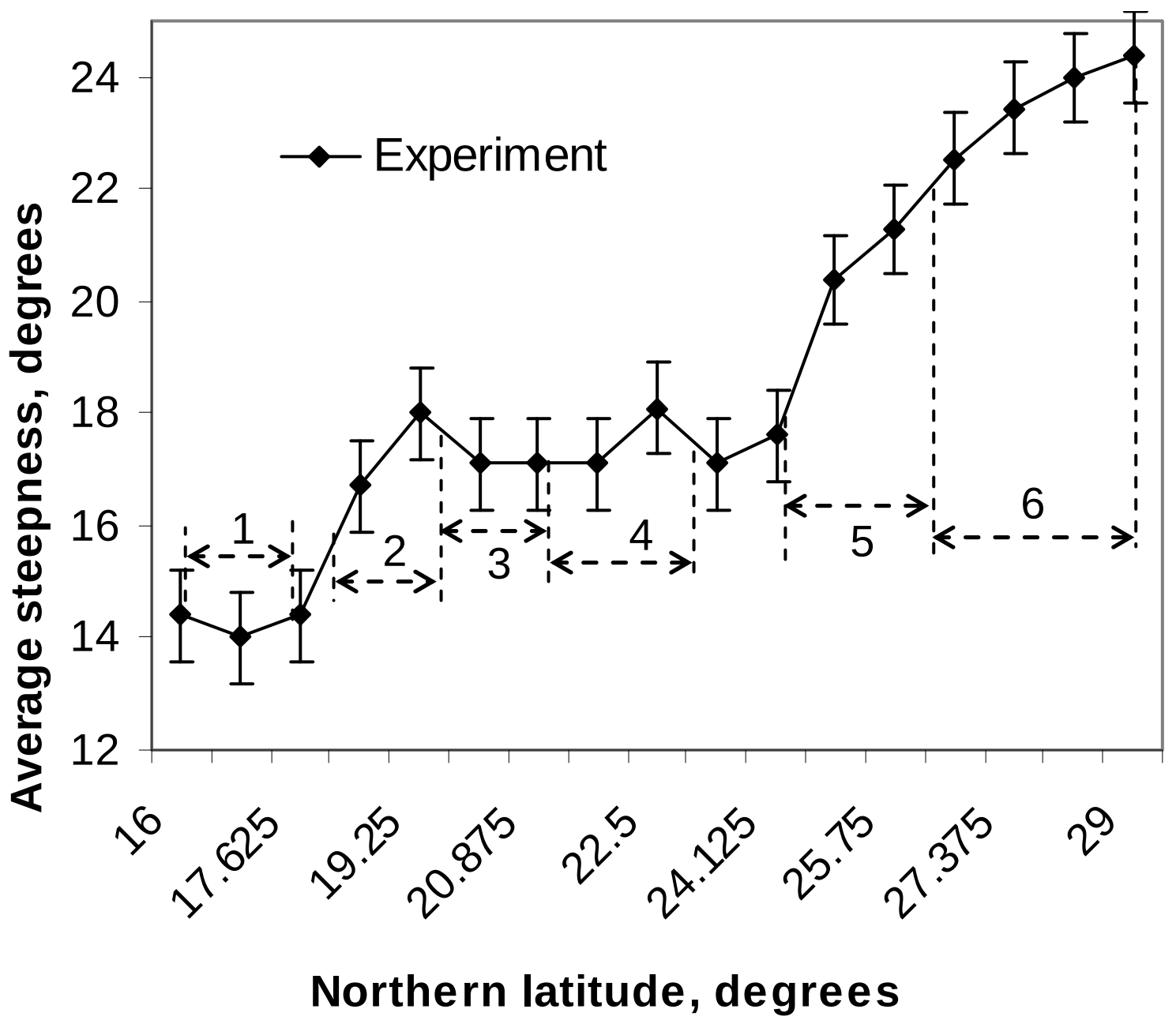

Fig. 4. Change of average roughness of the surface along the observed trajectory based on experimental data.

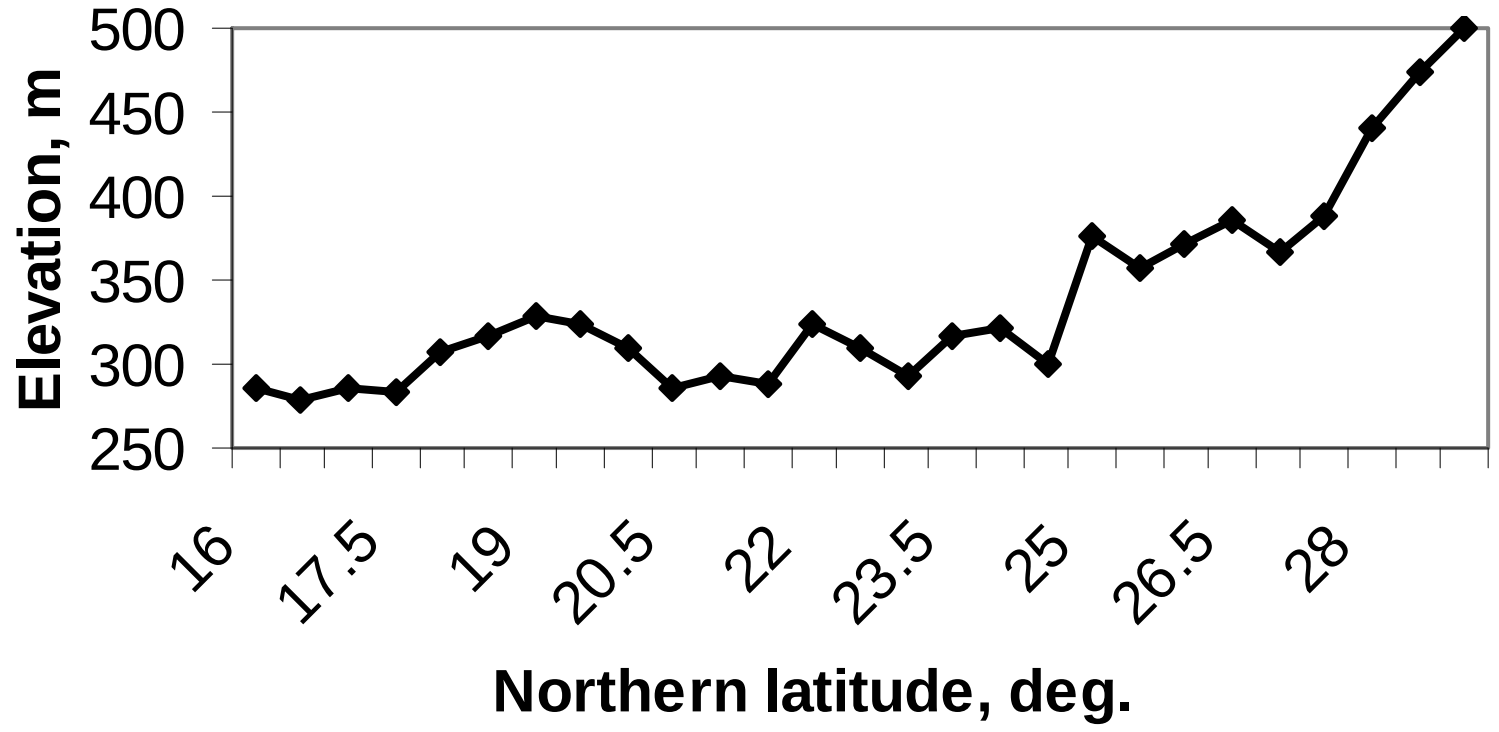

Fig. 5. Elevation profile.

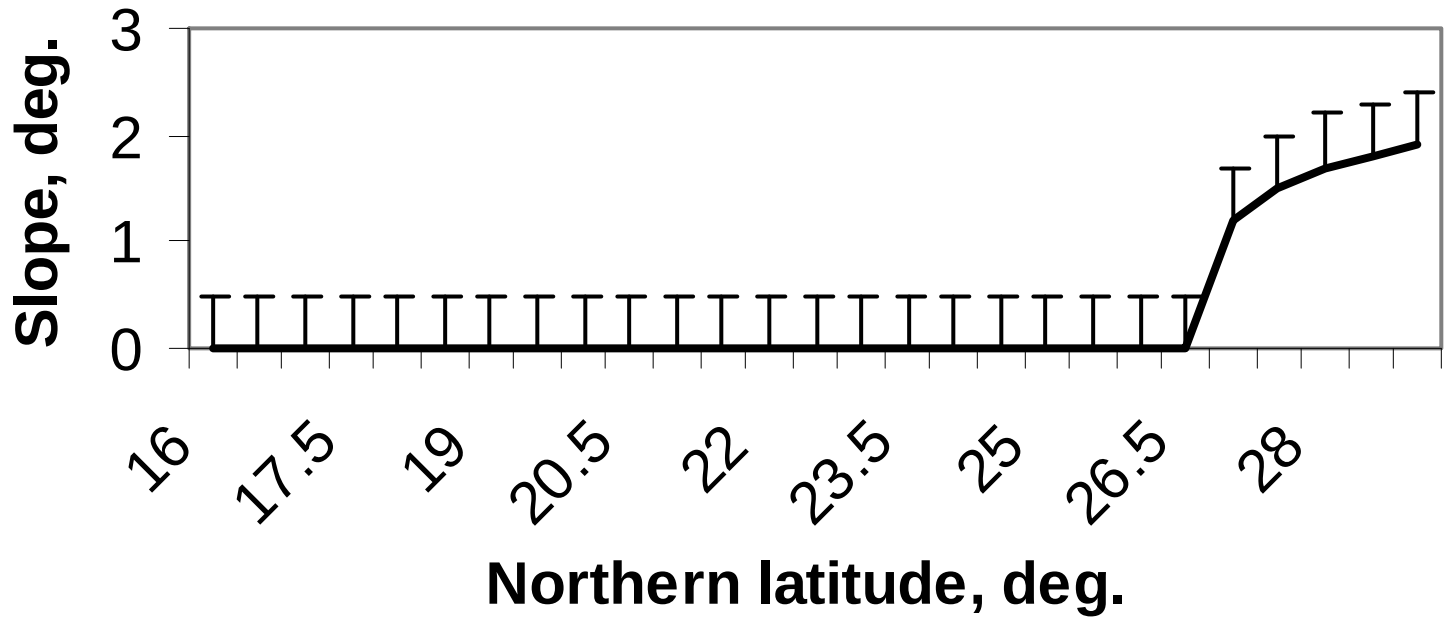

Fig. 6. Geographical slope.

In Fig. 4, we identify the following geographical areas.

1. Flat area, south-east of Nema.

2. Plateau, sand dunes.

3. Jafene (sand, salt pans).

4. Erg Chache (large ridge of sand dunes).

5. Gbat El Eglab (plateau).

6. Plateau, Atlas Mountains foothills.

For determining the average slope from the value of the roughness coefficient, we used statistical models of rough surfaces, including ones with variations in the slope value [9,17]. In these models, the presence of slope variations leads to curves which are positioned slightly lower than the curves for a rough surface with a constant slope.

Therefore, the discussed results confirm validity of the theoretical study. Overall, at observation angles of more than thirty degrees, the roughness coefficient unambiguously correlates with geometry of surface irregularities, defined as the average slope of surface roughness. This property of the roughness coefficient can be used in different applications. For instance, the roughness coefficient can be used to determine geometrical characteristics of surfaces in remote sensing and radar observations or for recognition purposes. It can also be used to measure the surface roughness of items in production environments by using, for instance, laser radiation. In the last case, the measurement system can be calibrated by sample items with known roughness or by statistical evaluations, similar to approaches described in [8].

Another advantageous feature of introduced roughness coefficient is that it depends less on the dielectric permittivity than the geometry of surface irregularities. Combined with the fact that the dielectric permittivity can be accurately estimated using many methods, it means that the roughness coefficient can provide an accurate characterization of surface's geometrical characteristics.

On the other hand, if we know the roughness characteristics, then we can interpret polarization measurements more accurately. For instance, if we know the surface's roughness coefficient in terms of emissivities, then we can rewrite (16) as follows.

$$T = \frac{T_s^2}{S_\chi (2T_s - T_p)} \tag{28}$$

This way, we can more accurately determine the temperature of a rough surface.

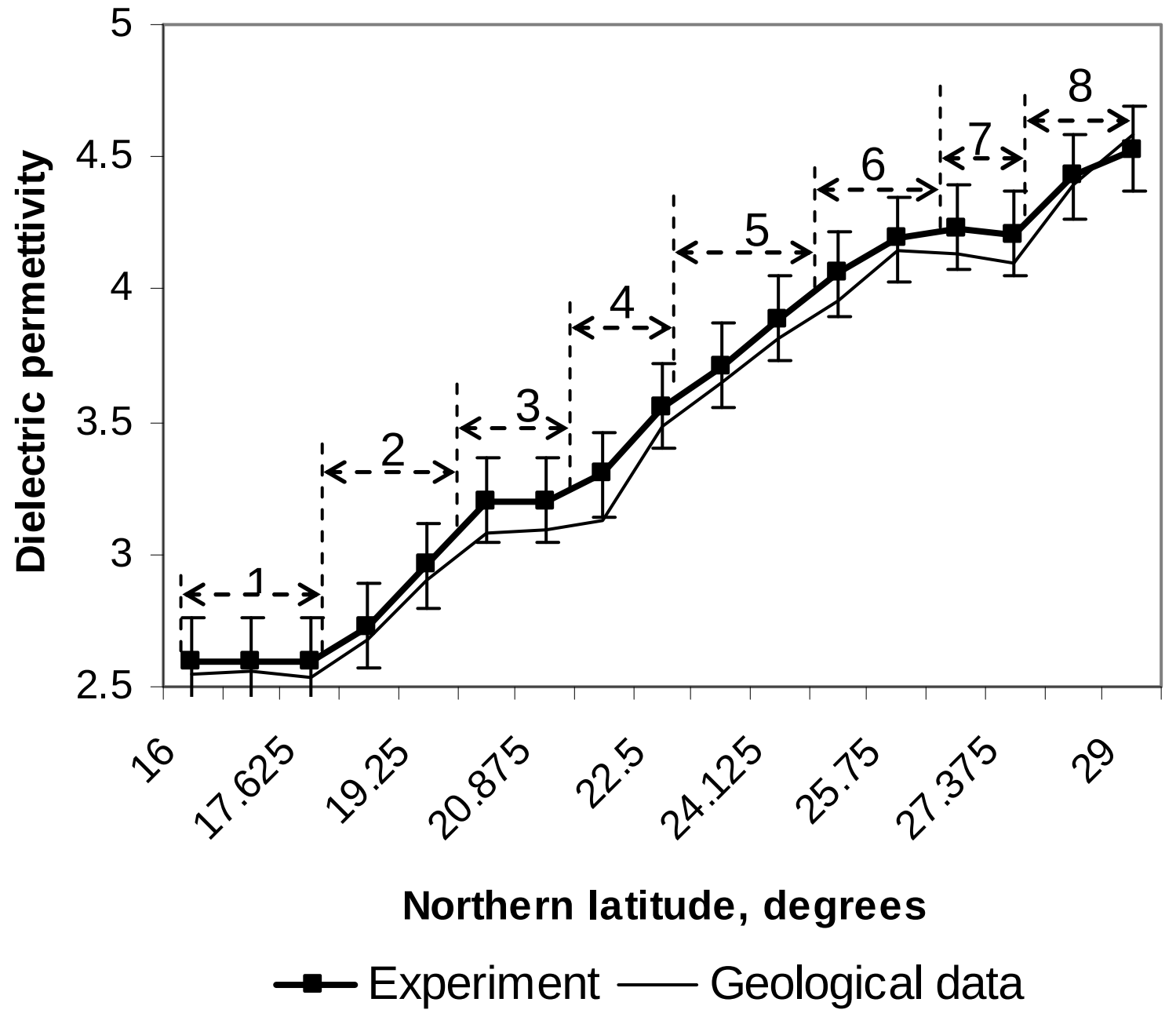

Fig. 7. Change of dielectric permittivity along the observed trajectory.

Fig. 7 shows the change of dielectric permittivity along the trajectory of antenna spot. We found the following zones, shown in Fig. 7, for which we can identify the types of minerals, rocks and soils, and the appropriate values of dielectric permittivity.

1. Sandstone, sand patches (fluisol).

2. Sand, soil (alisols).

3. Sand dunes (sand from rock formations), salt pans (arenosols).

4. Sand dunes ridge (Erg Chache) (alisols).

5. Salt pans, sand, rock formations (alisols).

6. Ridges of sand dunes on plateau El Glab. Granite and pre-Cambrian era metamorphic rocks, such as shale, silts (protrusion of a continental platform). Alisosols, leptosols (thin, shallow soils on hard rock).
7. Rock formations on plateau, with ridges of sand.
8. Atlas Mountains foothills. Rock formations (granite, metamorphic rocks), patches of sand.

We then used optical images in order to identify the minerals in particular areas and to determine their surface coverage. Values of dielectric permittivity for particular minerals at comparable frequencies are available from different sources, in particular [11,12]. We also compared these values of dielectric permittivity with data previously collected from other sources, including unpublished reports, and our direct measurements of dielectric permittivity in the range of 0.8-8.5 cm in the laboratory settings, and found the results of this comparison satisfactory.

The dependence between the emissivity and the dielectric permittivity of emitting surface is non-linear. We used the following approach to calculate dielectric permittivity on the basis of geological data and optical observations. The integral emissivity $\chi_T$ of the surface that is composed of several types of minerals, each having emissivity $\chi_i$ and the surface coverage weight $W_i$, is defined as $\chi_T = \sum_{i=1}^{i=N} W_i \chi_i$. Using (1) at $\alpha = 0$ and the relationship $\chi_{T(\alpha=0)} = 1 - \left| \frac{\sqrt{\varepsilon_T} - 1}{\sqrt{\varepsilon_T} + 1} \right|^2$, we can transform this equation to $\left|\frac{\sqrt{\varepsilon_T}-1}{\sqrt{\varepsilon_T}+1}\right|^2 = E$ , where $E = \sum_{i=1}^{i=N} W_i \left|\frac{\sqrt{\varepsilon_i}-1}{\sqrt{\varepsilon_i}+1}\right|^2$ . So, we can find the effective dielectric permittivity $\varepsilon_T$ for a surface that includes several emitting objects with different dielectric permittivities as $\varepsilon_T = \left(\frac{1+\sqrt{E}}{1-\sqrt{E}}\right)^2$ . Thus obtained results were in very good agreement with the values of dielectric permittivity found from satellite measurements on the basis of (9). Although additional research is needed, the suggested approach demonstrates high potential for practical applications.

## Conclusion

The results of this study showed that the combined measurements of electromagnetic radiation using orthogonal linear polarizations allow retrieving more information about the characteristics of reflected and emitted signals, such as signals' amplitudes and phase structure, and also allow finding a medium's parameters (dielectric permittivity, temperature, average surface slopes).

We also considered certain relationships between reflection coefficients that hold true for all angles of observation, which become especially simple when the angle of observation is $45^0$ . Similar polarization invariants exist for emissivities. It turns out that such introduced polarization invariants have a practical value. They can be used for evaluation of geometry, temperature and dielectric permittivity of observed surface, calibration of receiving and emitting devices and evaluation of external noise. Experimental observations confirmed the validity and practical value of proposed methods.

On a fundamental physical level, there is a certain geometrical symmetry in the interaction of electrical dipoles, located on and near the medium's surface, with linearly polarized electromagnetic waves incident at an angle of forty five degrees. This effect may serve as one of the connection points between geometrical and physical optics and theory of electromagnetism, which would also allow taking into account finer effects, such as, for instance, the aforementioned finite thickness of medium's layer in which the refraction and reflection processes take place.

## Acknowledgement

The author greatly appreciates the efforts of Alexander Shestopaloff, whose dedicated and enthusiastic help in retrieving geographical, climate, soil and mineral related information was crucial for this project. The author thanks A. Kyriakos for the discussion and advices.

## Appendix

Let us find $\sqrt{\varepsilon - \sin^2\alpha}$ and $\varepsilon$ from (1).

$$\sqrt{\varepsilon - \sin^2\alpha} = \frac{\cos\alpha(1+R_s)}{1-R_s}. \qquad \text{(A1)}$$

$$\varepsilon = \sin^2\alpha + \cos^2\alpha\left(\frac{1+R_s}{1-R_s}\right)^2 = \frac{1+R_s^2+2R_s\cos 2\alpha}{(1-R_s)^2}. \qquad \text{(A2)}$$

Substituting these values into (2) and doing appropriate transformations, we obtain the following.

$$R_p = \frac{\cos\alpha \dfrac{1+R_s^2+2R_s\cos 2\alpha}{(1-R_s)^2} - \dfrac{\cos\alpha(1+R_s)}{1-R_s}}{\cos\alpha \dfrac{1+R_s^2+2R_s\cos 2\alpha}{(1-R_s)^2} + \dfrac{\cos\alpha(1+R_s)}{1-R_s}} = \tag{A3}$$
$$= \frac{\cos\alpha(1+R_s^2+2R_s\cos 2\alpha)-\cos\alpha(1-R_s^2)}{\cos\alpha(1+R_s^2+2R_s\cos 2\alpha)+\cos\alpha(1-R_s^2)} = \frac{R_s^2+R_s\cos 2\alpha}{1+R_s\cos 2\alpha}.$$

If we substitute $R_s\cos 2\alpha$ from (A3) into (A2) and perform appropriate transformations, we obtain the following.

$$\varepsilon = \frac{1+R_s^2+2R_s\cos 2\alpha}{(1-R_s)^2} = \frac{1+R_s^2+2\dfrac{(R_p-R_s^2)}{(1-R_p)}}{(1-R_s)^2} = \frac{(1+R_p)(1+R_s)}{(1-R_p)(1-R_s)}. \tag{A4}$$

Below, we consider the case when magnetic permittivities of media differ from one. Let us denote $\mu = \mu_1/\mu_2$. Indexes “1” and “2” indicate accordingly the medium from which the incident wave comes, and the reflecting medium. Using general formulas for reflection coefficients from [7], we can rewrite (1) and (2) as follows.

$$R_s = \frac{\mu\sqrt{\varepsilon-\sin^2\alpha}-\cos\alpha}{\mu\sqrt{\varepsilon-\sin^2\alpha}+\cos\alpha}. \tag{A5}$$

$$R_p = \frac{\mu\varepsilon\cos\alpha-\sqrt{\varepsilon-\sin^2\alpha}}{\mu\varepsilon\cos\alpha+\sqrt{\varepsilon-\sin^2\alpha}}. \tag{A6}$$

Here, $\varepsilon = \varepsilon_2/\varepsilon_1$, the same as in (1) and (2). Doing transformations of (A5) and (A6) similar to (A1) – (A3), we obtain.

$$R_p = 1-\frac{2(1-R_s^2)}{s^2\mu^2(1-R_s)^2+c^2(1+R_s)^2+1-R_s^2}. \tag{A7}$$

When $\mu = 1$, (A7) transforms to (A3). When $\alpha = 45^0$, (A7) transforms to equation:

$$R_p = 1 - \frac{4(1 - R_s^2)}{\mu^2 (1 - R_s)^2 + (1 + R_s)^2 + 2(1 - R_s^2)}. \qquad \text{(A8)}$$

If we assume $\mu = 1$, then (A8) transforms to (7), that is $R_p = R_s^2$.

When $\varepsilon = 1$, that is $\varepsilon_1 = \varepsilon_2$, then, as it follows from (A5) and (A6), $R_s = R_p = (\mu - 1)/(\mu + 1)$. In other words, the reflection coefficients are the same on both polarizations and do not depend on the observation angle. Creating different substances with the same dielectric permittivity is possible; for instance, using mixtures or solutions. In this case, the reflection and refraction properties of the boundary between such substances will depend only on the magnetic permittivity, which in some cases can be useful.